\documentclass[prx,twocolumn,english]{revtex4-1}
\usepackage{amsmath,amssymb,mathrsfs}
\usepackage{natbib}
\usepackage{subfigure}
\usepackage{tabularx}
\usepackage{epsfig}
\usepackage{longtable}
\usepackage{amsfonts}
\usepackage{rotating}
\usepackage{subfigure}
\usepackage{amsmath}
\usepackage{babel}
\usepackage{comment}
\usepackage{bbold} 

\usepackage[unicode=true,
 bookmarks=true,bookmarksnumbered=false,bookmarksopen=false,
 breaklinks=false,pdfborder={0 0 1},backref=false,colorlinks=true]
 {hyperref}
\hypersetup{
linkcolor=magenta, urlcolor=blue, citecolor=blue, pdfstartview={FitH}, hyperfootnotes=false, unicode=true}

\usepackage{color}

\newcommand{\bea}{\begin{eqnarray}}
\newcommand{\eea}{\end{eqnarray}}

\newcommand{\be}{\begin{equation}}
\newcommand{\ee}{\end{equation}}



\begin{document}
\title{Statistical Bubble Localization with Random Interactions} 
\author{Xiaopeng Li}
\author{Dong-Ling Deng} 
\author{Yang-Le Wu}
\author{S. Das Sarma} 
\affiliation{Condensed Matter Theory Center and Joint Quantum Institute, Department
of Physics, University of Maryland, College Park, MD 20742-4111, USA}

\begin{abstract}
We study one-dimensional spinless fermions with random interactions, but {\it without} any on-site disorder.
We find that random interactions generically stabilize a many-body 
localized phase, in spite of the completely extended single-particle degrees of freedom.
In the large randomness limit, we construct
``bubble-neck'' eigenstates having a universal area-law entanglement entropy on average, with the number of volume-law states being  exponentially 
suppressed. We argue that this statistical localization is beyond the 
phenomenological local-integrals-of-motion description of many-body 
localization.  With exact diagonalization, we confirm the robustness of the 
many-body localized phase at finite randomness by investigating 
eigenstate properties such as level statistics, entanglement/participation 
entropies, and nonergodic quantum dynamics. At weak random interactions, the 
system develops a thermalization transition when the single-particle hopping  
becomes dominant.  
\end{abstract}
\maketitle

Disorder in isolated quantum systems leads to  fascinating phenomena such 
as Anderson localization~\cite{AndersonLocalization}.  Non-interacting particles in the Anderson localized phase form a perfect 
insulator with vanishing DC conductivity even at infinite 
temperature.  The lack of thermal transport in an Anderson localized system 
prohibits thermalization, making it intrinsically nonergodic and far out of 
equilibrium.
The stability of localization and non-ergodicity against 
interactions, however, remained controversial until the recent study of 
many-body localization (MBL)~\cite{Basko06,huse2015many,2015_Altman_review}.  
Following the perturbative analysis in Ref.~\cite{Basko06},
the robustness of localization against interactions
has now been established through exact numerical
calculations~\cite{oganesyan2007,pal2010,moore2012,2013_Bauer_JSM,vadim,bardarson2014,2015_Singh_MBL_PRL,2015_Khemani_Pollmann_PRL,2015_Yu_Pekker_arXiv,2015_Lim_Sheng_PRB,2016_Kennes_Karrasch_PRB}
and a mathematical proof under certain reasonable 
assumptions~\cite{imbrie2014many}.
Experimentally, the dynamical nonergodic 
aspects of the MBL phase have been examined with cold atoms in optical 
lattices~\cite{blochmbl,2016_Bordia_Bloch_MBL_PRL,2016_Bordia_Bloch_arXiv,2016_Choi_Bloch_MBL_Science} and trapped ions~\cite{2015_Monroe_MBL_NatPhys}.  
Although currently an active area of research, the general consensus is that a noninteracting quantum system with sufficiently strong single-particle (i.e. on-site) disorder remains many-body-localized in the presence of finite interparticle interactions.

While the existence of MBL is accepted for interacting 
disordered fermions, the role of interaction remains somewhat tangential.
In the numerical studies of models with on-site disorder, MBL is only found in 
the regime dominated by single-particle disorder potentials where the noninteracting system is necessarily strongly localized~\cite{oganesyan2007,pal2010,moore2012,2013_Bauer_JSM,vadim,bardarson2014,2015_Singh_MBL_PRL}.
Mathematically, despite the proof of existence of MBL~\cite{imbrie2014many}, a 
lower bound for the required disorder strength has not been established.
In the ``local-integrals-of-motion'' (LIM) 
description~\cite{2014_Huse_MBL_PRB,2013_Serbyn_PRL,chandran2015,Ros2015420}, 
the conserved charges strongly resemble their non-interacting counterparts in 
the deep MBL regime.
It is difficult to single out the effect of interaction for MBL in 
models with single-particle disorder, where interaction and single-particle 
terms are always intertwined.
This issue is particularly worrisome when one looks for ``smoking-gun''
experimental signatures to distinguish MBL from Anderson localization, and the possibility that all experimentally observed MBL phenomena are essentially (slightly perturbed) single-particle Anderson localization cannot be definitively ruled out.
It is thus desirable to study a simpler system where the localization is 
driven purely by many-body effects, 
and the interacting MBL phase is not adiabatically connected to a single-particle Anderson localized phase.

\begin{figure}[htp] 
\includegraphics[angle=0,width=\linewidth]{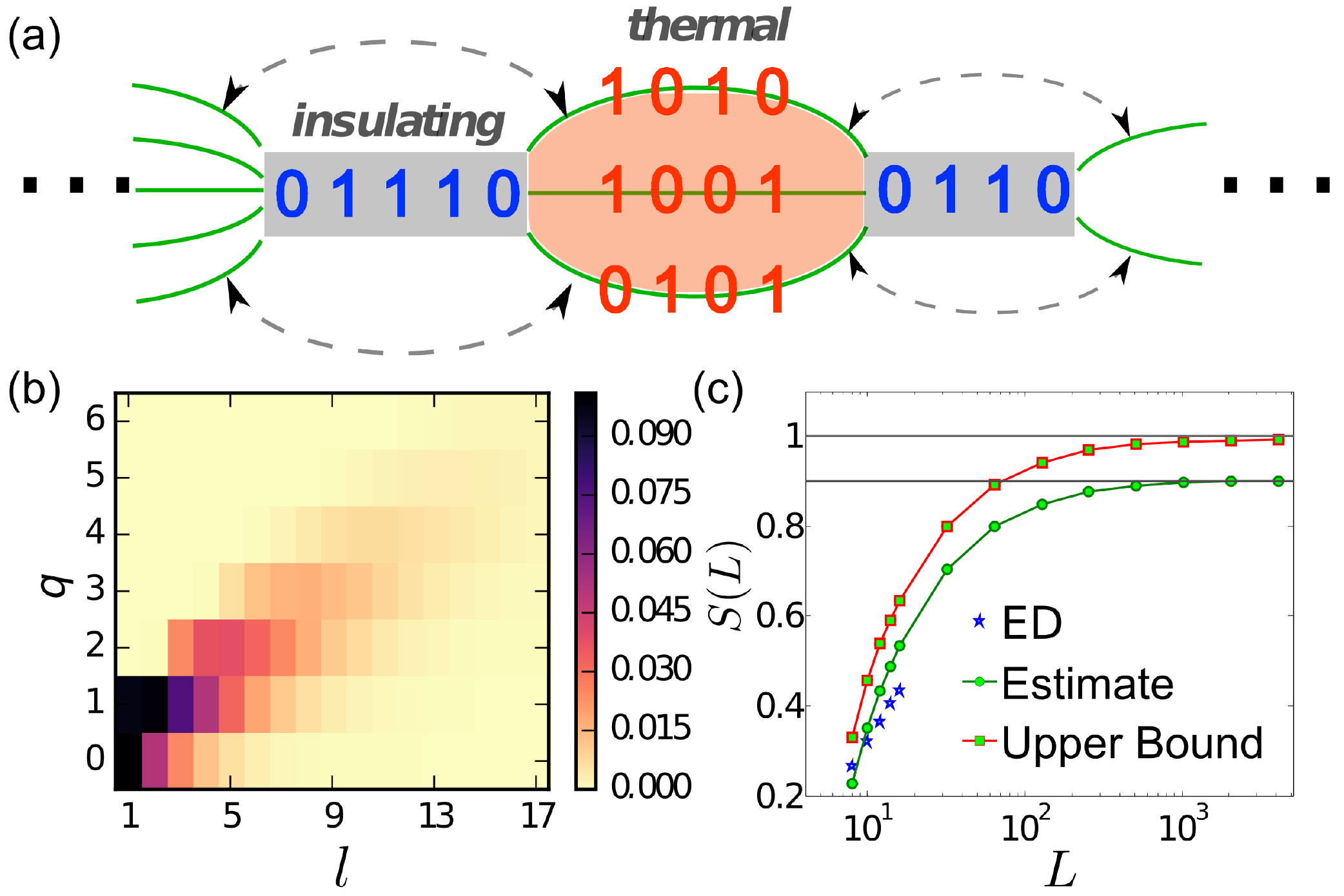}
\caption{Eigenstates in the infinite random interaction limit. (a) illustrates the ``bubble-neck'' eigenstates.  Clusters with more than one 
fermion on adjacent sites form insulating blocks (IB). Others with isolated fermions 
form thermal bubbles (TB). Quantum superpositions are allowed (forbidden) in 
the TB (IB). Cross-block tunnelings (dashed arrows) are negligible in this 
limit.  (b) shows the histogram of 
the size ($l$) and the particle number ($q$) of TBs. (c) shows 
the bipartite entanglement entropy. The
symbols `$\star$' correspond to  exact 
diagonalization results with $W/t = 10^3$. The solid lines correspond to the 
entanglement upper bound and the Page-value estimate from random sampling, 
which respectively saturate to $S_{\rm ub} \approx 1$ and 
$S_{\rm est} \approx 0.9$ in the $L\to \infty$ limit.
 }
\label{fig:Fig0} 
\end{figure}

In this paper, we consider the precise opposite limit 
and study MBL in a random-interaction model, whose 
non-interacting limit is completely extended.  In the strong 
randomness limit, we formulate a ``bubble-neck'' construction (see 
Fig.~\ref{fig:Fig0}) for the MBL eigenstates in this system. Such bubble-neck eigenstates could have 
volume-law entanglement. 
Our construction hence goes beyond the scope of the LIM description and describes a novel type of MBL with no non-interacting analogue 
whatsoever (i.e. the corresponding noninteracting system is in a trivial extended phase). 
Further, we show that the average entanglement entropy over all such eigenstates still obeys an area law, and we provide a generic entropy upper bound, independent of the specific model realization of thermal bubbles. 
With exact 
numeric calculations, we confirm the robustness of the MBL phase at finite random 
interactions. For weak disorder, the system develops a thermalization 
transition when the single-particle tunneling effects become dominant overwhelming random interaction effects.  We 
stress that our proposed statistical bubble MBL phase is driven solely by the 
interaction, without any influence from single-particle on-site disorder.  
While aspects of MBL in the presence of extended single-particle orbitals 
have been discussed in other systems~\cite{2015_Li_MBL_PRL,modak2015many,2016_Li_Pixley_MBL_PRB,2016_Bauer_arXiv,2016_Romain_Potter_PRB,2016_Lev_arXiv}, our work shows that clean interacting spinless fermions 
have novel generic features distinct from previous studies, 
establishing that MBL in clean random interacting fermion systems is a generic phenomenon completely distinct from the MBL physics in disordered interacting systems which are adiabatically connected to Anderson localized systems as the interaction is turned off.

\textit{Model.---}We study one-dimensional (1D) spinless fermions with random nearest neighbor interactions, 
\be 
 H = -t \sum_{j=1} ^L  \left[ c_j ^\dag c_{j+1} + H.c. \right] + \sum_j V_j n_j n_{j+1}, 
\label{eq:Ham}
\ee 
where $c_j$ is a fermonic annihilation operator, $n_j = c_j ^\dag c_j $, $L$ 
is the number of lattice sites, and the tunneling $t$ is the energy unit 
throughout this paper.
We consider a uniform distribution for the random interactions 
$V_j \in [-W, W]$ and focus on half-filling.  In this model, the 
disorder effects arise purely from interactions, with the non-interacting degrees of 
freedom being completely delocalized.

\begin{figure}[htp] 
\includegraphics[angle=0,width=\linewidth]{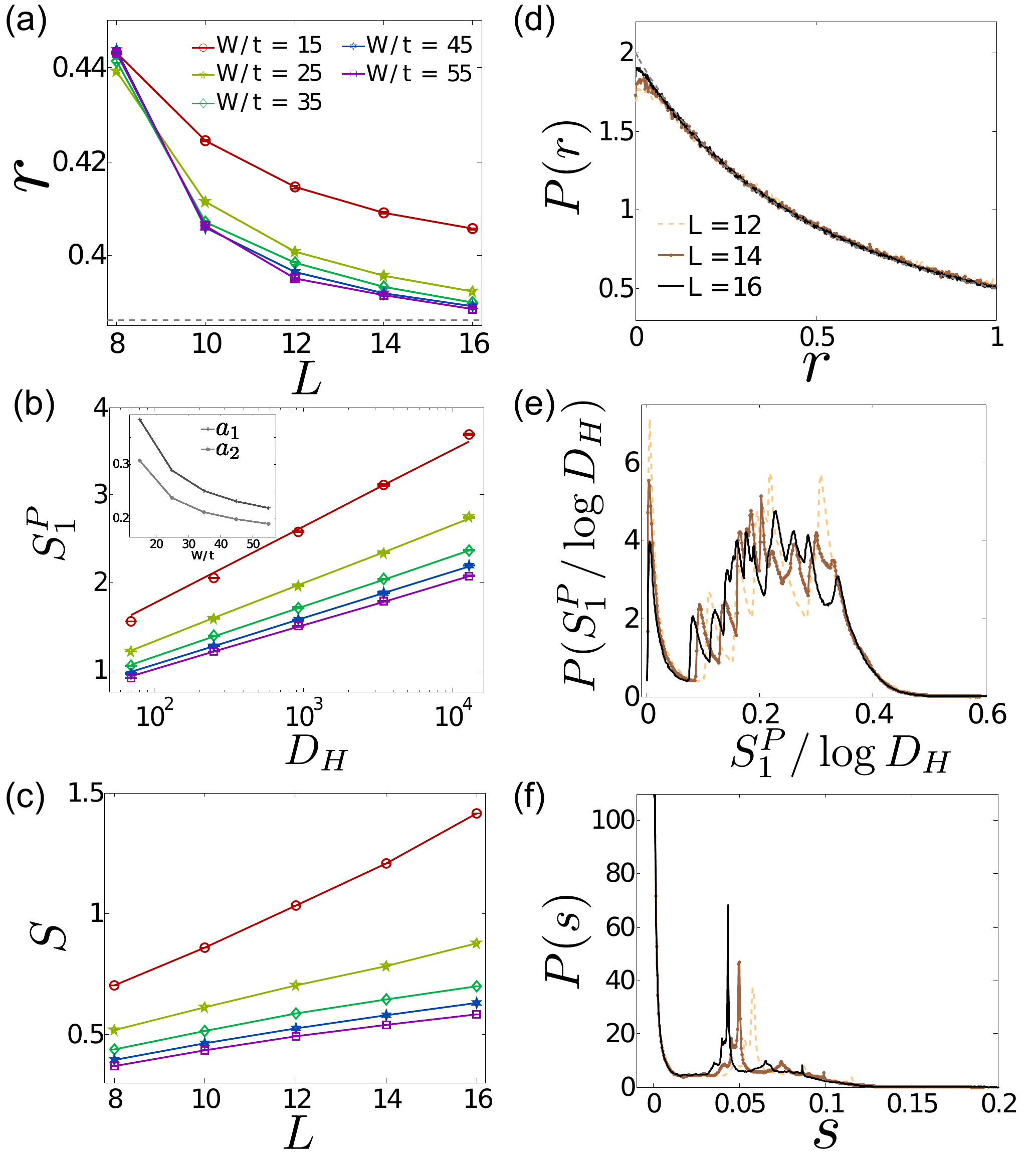}
 \caption{The MBL phase of the random interaction model (Eq.~\eqref{eq:Ham}). (a), (b), and (c) share the same legend and show the system-size dependence of $r$, $S_1^P$, and $S$, respectively. Their probability distributions across disorder samples and different eigenstates in the deep localized phase ($W/t = 55$) are shown in  (d), (e), and (f).  The dashed line in (a) marks the Poisson value $r_P = 2\log 2 -1$. In (d), the numeric data for $P(r)$ with $W/t = 55$ for different $L$ collapse to $P_0(r) = 2/(1+r)^2$ (grey dashed line) with the deviation barely noticeable in this plot.  The inset in (b) shows the participation entropy coefficients (see main text).
 } 
 \label{fig:Fig2}  
\end{figure}

{\it Analysis of the infinite randomness limit.---}
Let us first consider the strong randomness limit $W\to \infty$.
If the tunneling $t$ is strictly zero, the eigenstates of the system are trivial product states albeit with huge degeneracies. Turning on an infinitesimal tunneling breaks the degeneracy and gives a bubble-neck structure to the eigenstates to be described below.  

With infinitesimal tunneling (to the leading order in $t/W$), 
a cluster with more than 
one particles on adjacent sites (Fig.~\ref{fig:Fig0}) is localized (i.e., does 
not tunnel) due to random two-body interactions, and such clusters form {\it insulating blocks}.  Other clusters 
with isolated fermions are extended, forming {\it thermal bubbles}.  Fermions 
in the thermal bubbles can tunnel almost freely, except that the 
configurations with two fermions coming to adjacent sites are forbidden. 
A thermal bubble with $l$ lattice sites 
and $q$ fermions has a Hilbert space dimension 
$
D_{\rm therm} (l, q) = {l+1-q \choose q}.  
$ 
Fermion tunneling in a thermal bubble makes a finite many-body energy 
splitting of the order of $t/D_{\rm therm}$, 
which prohibits couplings of different thermal bubbles across insulating blocks (to leading order in $t/W$).  
The resulting bubble-neck eigenstates are illustrated in 
Fig.~\ref{fig:Fig0}. 
In the infinite randomness limit, only the thermal bubbles contribute to the 
entanglement entropy.
With random state sampling~\cite{citesupplement}, we find that the 
probability distribution of the thermal-bubble-size $P(l, q)$  decays 
exponentially for large $l$ (Fig.~\ref{fig:Fig0}b).
The entanglement entropy 
of the eigenstates in the large randomness limit is thus bounded, i.e., 
obeying an area-law scaling, which implies that the system is many-body 
localized (see Fig.~\ref{fig:Fig0}c for the explicit entanglement scaling). We 
find that the area-law entanglement entropy of such bubble-neck eigenstates 
has a {\it generic}  
upper bound with the Page-value 
estimate~\cite{1993_Page_PRL} in 
the thermodynamic limit, 
\bea
&& \textstyle S_{\rm ub} (L \to \infty) \approx 1, \\
&& \textstyle S_{\rm est} (L \to \infty) \approx 0.9,    
\eea 
independent of the specific model of thermal bubbles. Here the Page-value is the entanglement entropy averaged over random pure states~\cite{1993_Page_PRL}, and it provides an estimate for the entanglement in thermal states~\cite{2016_Sheng_Huse_arXiv}. 
The Page-value estimate agrees with our numeric exact diagonalization results for small systems (Fig.~\ref{fig:Fig0}c). 
We emphasize that the MBL eigenstates in the infinite interaction disorder limit are 
generic, independent of the specific disorder realizations.
The bubble-neck MBL picture with generic 
statistical entanglement properties does not depend on the specific 
model of the dynamics in the thermal bubble.

We stress that our MBL phase goes beyond the LIM description. In the 
LIM picture~\cite{2014_Huse_MBL_PRB,2013_Serbyn_PRL,chandran2015,Ros2015420}, all eigenstates for a fixed disorder configuration are 
short-range entangled with their entanglement entropy determined by certain 
localization length. In sharp contrast, the generic bubble-neck eigenstates 
(Fig.~\ref{fig:Fig0}a) could be long-range (volume-law) entangled although the 
number of such states is statistically suppressed 
by the exponentially decaying probability of long bubbles (Fig.~\ref{fig:Fig0}(b)). 
We thus conclude that our proposed random interaction driven MBL phase is 
sharply {\it distinct} from the on-site disorder driven MBL.

It is worth noting that the thermal bubble of the particular model in Eq.~\eqref{eq:Ham} is actually integrable through an inflated-fermion mapping approach~\cite{citesupplement}. However we stress that the physics presented here does not rely on the choice of this particular model.  We check this by  replacing the single-particle Hamiltonian with the Aubry-Andr\'e model where the thermal bubble is no longer integrable, finding quantitatively similar results~\cite{citesupplement}.


\begin{figure} 
\includegraphics[angle=0,width=\linewidth]{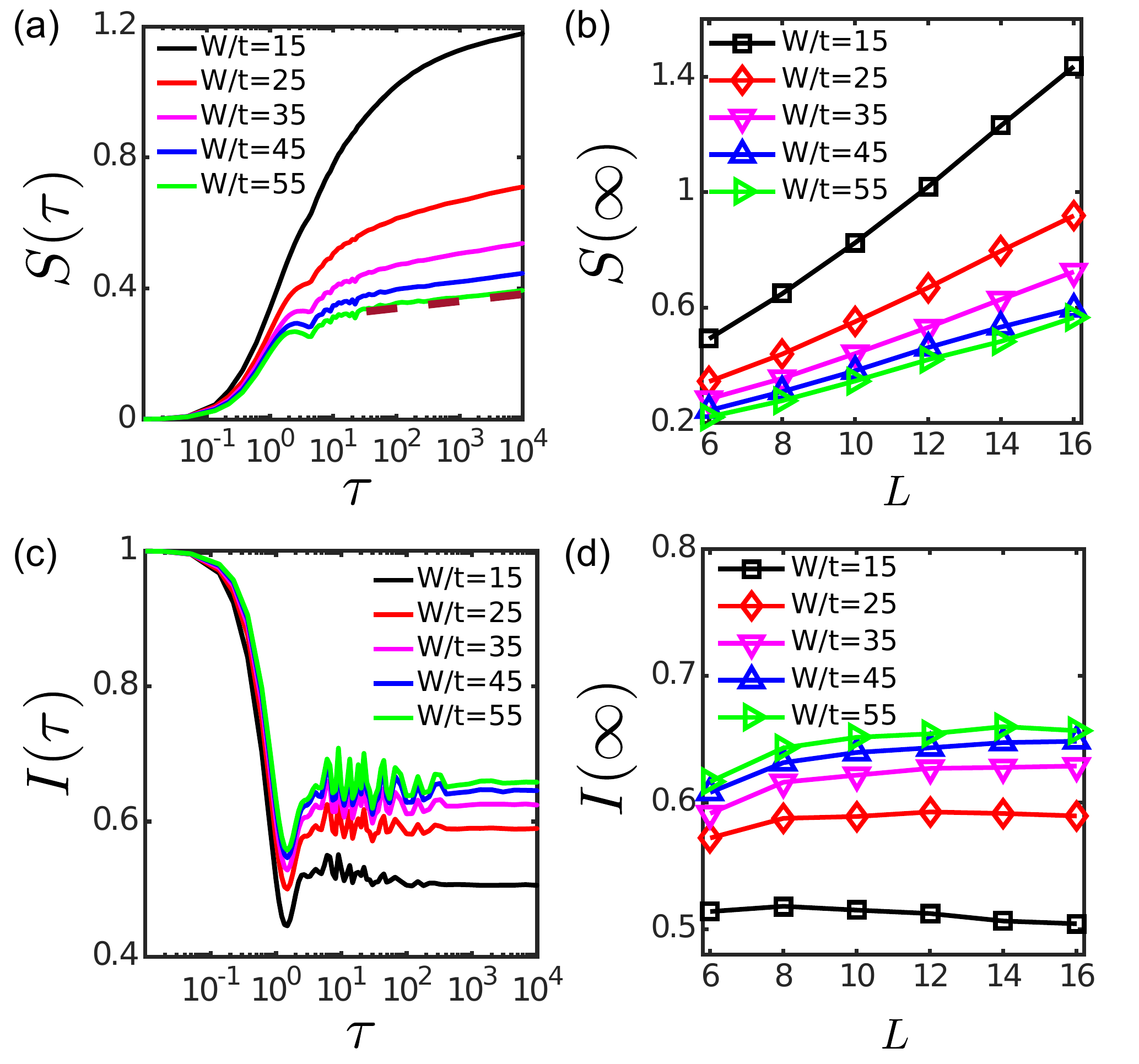}
 \caption{Dynamical properties of the localized phase. Here we begin with random product states and then compute their time-evolution with $H$ defined in Eq. (\ref{eq:Ham}).  (a) Entanglement growth with varying $W/t$. The dashed brown line is a logarithmic fit to the data.  (b) Scaling of the saturated value of entanglement $S(\infty)$. Similar to the case of random on-site disorder driven MBL~\cite{moore2012}, for strong random interaction with $W >25$, the entanglement entropy exhibits an unbounded logarithmic growth in the thermodynamic  limit and the its saturation value obeys a volume-law for finite $L$. This lends strong support of a random interaction driven MBL phase. (c)Dynamical evolution of density imbalance. It saturates to a finite value and thus does not relax at long-time, another signature of MBL and ergodicity breaking. (d) Density imbalance at long time limit as a function of $L$ for different $W/t$.} 
 \label{fig:Fig5}  
\end{figure}

\textit{The MBL phase at finite randomness.---}With finite random interaction, 
the ``forbidden'' cross-block couplings (Fig.~\ref{fig:Fig0}a) come into play 
and our bubble-neck picture no longer strictly applies. 
We study such effects using exact diagonalization. We have 
investigated different diagnostics, the bipartite entanglement entropy ($S$), 
the level statistics gap ratio ($r$), and the wave-function participation  
entropy ($S_m ^p$), which are widely used in the literature to characterize 
MBL. The entanglement entropy $S$ signifies localization in real space. The 
gap ratio that characterizes the level statistics is  defined to be 
$r \equiv {\rm min}(\delta_n, \delta_{n+1} ) /{\rm max} (\delta_n, \delta_{n+1})$~\cite{oganesyan2007}, with $\delta_n$  the energy spacing between close-by eigenstates. The participation entropy~\cite{1972_Bell_RPP,1980_Wegner_ZPB,2011_Rodriguez_PRB,2014_Luitz_Alet_PRL} is introduced to quantify the localization property in the many-body Hilbert space, 
$
S_m ^P = \frac{1}{1-m} \sum_{\{n\} } |\Psi_{\{n\}} | ^{2m} 
$, 
with $S_1 ^P =- \sum_{\{n\} }| \Psi_{\{n\}}|^2 \log |\Psi_{\{n\}}|^2 $, where  
$\Psi_{\{n\}}$ is the many-body wave function. 
We average over $1000$ ($10000$) disorder realizations for systems with size 
$L \ge 12$ ($L < 12$). Within each disorder realization, we average over all 
eigenstates with an equal weight, corresponding to an ``infinite temperature'' 
ensemble.   

In Fig.~\ref{fig:Fig2}, we provide the system-size dependence and the probability distributions of different quantities. Fig.~\ref{fig:Fig2}(a) shows the average gap ratio with varying random interaction strength $W/t$. This quantity approaches the GOE (Gaussian Orthogonal Ensemble) value $r_G \approx 0.53$ in the thermal phase and the Poisson value $r_P = 2\log 2 -1$ in the nonergodic MBL phase. 
At strong random interaction ($W/t \in [25, 55]$ shown in the figure) $r$ monotonically decreases as we increase the system size, and systematically  approaches the universal Poisson value $r_P$ in the thermodynamic limit (Fig.~\ref{fig:Fig2}(a)). Moreover, the probability distribution of the gap ratio for different eigenstates and disorder samples collapses to the function of $P_0 (r) = 2/(1+r)^2$ (Fig.~\ref{fig:Fig2}(d)), which corresponds to the precise Poisson level statistics. We attribute the small deviation from $P_0(r)$ to finite-size effects as it systematically shrinks on increasing $L$.

Fig.~\ref{fig:Fig2}(b) shows the rank-$1$ participation entropy  $S_1 ^P$.  In 
the thermal phase with its wave function completely delocalized in the Hilbert 
space, $S_1 ^P$ will approach $\log D_H$ ($D_H$ is the Hilbert space 
dimension) in the thermodynamic limit, whereas in the localized phase 
$S_1^P/\log D_H<1$ meaning the wave function does not spread over the entire 
Hilbert space. In our numerics, we find that $S_1^P$ is proportional to $\log D_H$, 
$S_1 ^P = a_1 \log D_H$, with the coefficient $a_1 \ll 1$ for $W/t \ge25$. (It 
is worth noting that a related quantity, normalized participation 
ratio~\cite{vadim}, decays exponentially with the system size.) This 
implies wave function localization in the Hilbert space. The broad 
distribution of $S_1 ^P$ (Fig.~\ref{fig:Fig2}d) indicates a large variance of 
dominant thermal bubble sizes in different eigenstates. We also calculated the 
rank-$2$ participation entropy and found its coefficient $a_2 \ll 1 $
($a_2 = S_2 ^P/\log D_H$), further verifying the localization of the system. It 
is worth mentioning that $a_2 \neq a_1$ (the inset of Fig.~\ref{fig:Fig2}(b)), 
indicating that this random interaction driven MBL phase is multi-fractal. The 
broad distribution of participation entropy $P(S_1^p)$ shown in 
Fig.~\ref{fig:Fig2} (e) is consistent with the multi-fractal behavior.

Fig.~\ref{fig:Fig2}(c) shows the bipartite entanglement entropy. We find that it 
grows with increasing $L$ even for very strong random interactions (we have 
checked the entanglement scaling for $W/t$ up to $10^6$). At the same time, 
$S(L)$ apparently bends downwards for $W/t \ge 35$. We attribute the growth of $S(L)$ to 
finite size effect,  as even at infinite randomness limit we still see  strong $L$ dependence in $S(L)$ for 
$L$ up to $100$ (Fig.~\ref{fig:Fig0}(c)).  
In the 
distribution $P(s)$ shown in Fig.~\ref{fig:Fig2}(f) we 
find $P(s \to 0)$ tends to diverge as $L$ increases. This signifies the robustness of 
insulating blocks for finite random interaction.


\textit{Entanglement dynamics and quantum non-ergodicity.---} 
To further verify the MBL phase, we study the quantum dynamics by initializing 
the system in random product states. The time-dependent entanglement entropy
($S(\tau)$) and number imbalance ($I(\tau)$) are monitored (Fig.~\ref{fig:Fig5}). The number 
imbalance is defined as
$$ 
 I (\tau) = \frac{N_{1} (\tau) - N_0(\tau)} {N_{1} (\tau) + N_0(\tau) }, 
$$ 
with $N_1$ ($N_0$) referring to number of particles in the initially occupied (unoccupied) lattice sites.  For the number imbalance (Fig.~\ref{fig:Fig5}(c)), we find that it does not relax at long time for large $W/t$, confirming the dynamical nonergodicity of the system. For $S(\tau)$ (Fig.~\ref{fig:Fig5}(a)), we obtain a linear growth at the beginning up to a ballistic time scale $\tau_0$, and  logarithmic growth at later time, which is qualitatively  similar to the case of on-site disorder driven MBL. But there are two quantitative differences from the on-site disorder case. One is that the ballistic time scale $\tau_0$ is about several tunneling time even at huge $W$.  We expect $\tau_0$  to be the tunneling time multiplied by the typical thermal-bubble size in our bubble-neck MBL phase. The other  is that the long time limit of entanglement entropy $S(\infty)$ is significantly larger than the deep on-site disorder MBL phase, which we attribute to the existence of thermal bubbles in our MBL system.

\begin{figure}[htp] 
\includegraphics[angle=0,width=\linewidth]{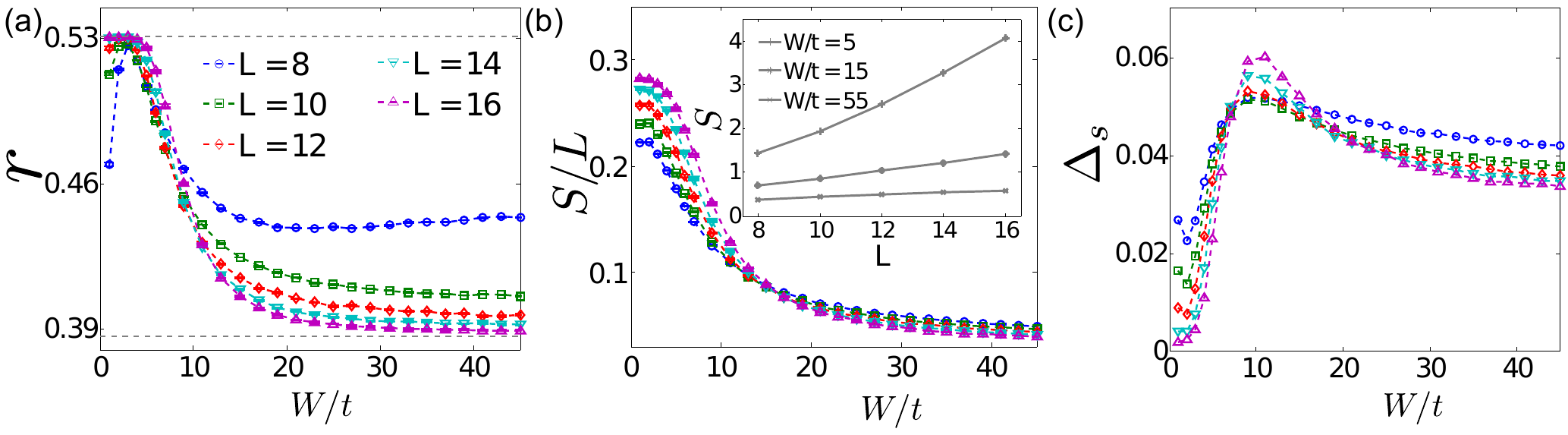}
 \caption{MBL transition of 1D fermions with random interactions.  (a) Disorder averaged adjacent gap ratio $r$ as a function of the random-interaction strength $W/t$. The level statistics obey the GOE and Poisson distributions  in the thermal and MBL phase with $r$ approaching $0.53$ and $2\log 2 -1$ (marked by `dashed' lines), respectively. The transition gets sharper as we increase the system size $L$. (b) Disorder averaged half-chain entanglement entropy density $S/L$. The inset shows the scaling of $S$ with increasing $L$. The entanglement entropy is strongly suppressed at large $W$.  (c) Standard deviation of $s$ across disorder samples ($\Delta_s$).  The numeric results indicate a MBL transition locating at $W_{\rm c} /t \in (5, 15)$.  }
\label{fig:Fig1} 
\end{figure}

\textit{The MBL transition at finite $W/t$.---}As we further decrease $W/t$, the cross-block couplings (Fig.~\ref{fig:Fig0} (a)) become more important and eventually drive a delocalization/thermalization transition. 
Fig.~\ref{fig:Fig1} shows the behavior of the different diagnostics. Fig.~\ref{fig:Fig1}(a) shows the gap ratio $r$. At 
strong randomness $W/t >20$, $r$ approximately stays at the universal Poisson  
value. For  $W/t < 5$, we find that $r$ systematically approaches the GOE 
value $r_G$ with increasing $L$, which implies that the system is in a thermal 
phase. We expect $r(W/t)$ to approach a step function in the thermodynamic 
limit, giving a sharp transition at certain critical random-interaction 
strength $W_c$. The crossings for different lines in Fig.~\ref{fig:Fig1}(a)  
indicate $W_c/t$ lies between $5$ and $15$. Fig.~\ref{fig:Fig1}(b) shows the bipartite 
entanglement entropy density ($s = S/L$). At small $W$, the entanglement 
entropy obeys volume-law scaling, and is expected to approach the thermal 
entropy ($\sim 0.35L$) for large enough $L$. We find that the entanglement entropy 
has a plateau-like behavior for small $W/t$, providing numerical evidence for 
$s$ to be a constant in the thermal phase. Fig.~\ref{fig:Fig1}(c) shows the 
variance of entanglement entropy ($\Delta_s$), which has been used to diagnose the MBL 
transition~\cite{bardarson2014,2015_Luitz_MBLedge_PRB,2015_Vosk_MBL_PRX,2016_Sheng_Huse_arXiv}. In calculating $\Delta_s$, we 
first average $s$ over  all eigenstates within one disorder realization, and 
then calculate the standard deviation across different samples. In our study of the random interaction model, we see $\Delta_s$ developing a peak in the crossover regime. The peak value grows significantly as we increase $L$, which is qualitatively similar to what has been found for the random on-site disorder models~\cite{bardarson2014,2015_Luitz_MBLedge_PRB,2016_Sheng_Huse_arXiv}.  This diverging behavior of the entanglement variance also suggests 
$W_c \in (5, 15)$. 

\textit{Conclusion.---} We study  random interaction driven MBL phase and point out its key differences with the on-site disorder driven case. We construct the generic bubble-neck eigenstates for the MBL phase in the infinite randomness limit, transcending the LIM description of MBL. With exact diagonalization, we confirm  the MBL phase at finite random interaction by calculating level statistics, participation entropy and entanglement dynamics. At weak random interaction, we find that the system undergoes a thermalization transition which is cross-block-tunneling-driven. 
The random interaction driven MBL discussed in this paper is generic for one-dimensional clean spinless fermions (as shown in~\cite{citesupplement} by studying different models) and is qualitatively different from MBL studied in interacting systems with single-particle disorder.


\paragraph*{Acknowledgment.---} 
This work is supported by JQI-NSF-PFC and LPS-MPO-CMTC. We acknowledge the University of Maryland supercomputing resources (http://www.it.umd.edu/hpcc) made available in conducting the research reported in this paper.

\newpage 

\begin{widetext}

{\centering \large \bf  
Statistical Bubble Localization with Random Interactions---Supplementary Materials}

\renewcommand{\thesection}{S-\arabic{section}}
\renewcommand{\theequation}{S\arabic{equation}}
\setcounter{equation}{0}  
\renewcommand{\thefigure}{S\arabic{figure}}
\setcounter{figure}{0}  


\section{Stochastic sampling of thermal bubbles} 
As shown in {\it Fig.~1 in the main text}, the eigenstates at infinite random 
interaction have a generic bubble-neck structure. In this section we discuss 
how to estimate the average entanglement entropy of such eigenstates with a 
stochastic sampling method. 

Note that only the bubbles that cross the two links between sites $L/2$ and $L/2+1$, 
and between sites $1$ and $L$ will contribute to the entanglement entropy
(we use the periodic boundary condition).
There are two different scenarios---(i) a single bubble spreads 
over both links, and (ii) two disconnected bubbles with one over each link.
The bubble configuration is then parametrized as 
$$ 
\alpha = (z, l_{k \in [1,z]}, q_k, l_k ^{\rm left} ),  
$$ 
with $z = 1,2$ representing the two different scenarios, $l_k$ and $q_k$ the size and particle number in each thermal bubble, $l_k ^{\rm left}$ the size of the thermal bubble within the left half of the system (with sites $j = 1, \ldots L/2$). Note that $\alpha$ can be thought as a function of either a Fock state or  an eigenstate.

The thermal bubble involves two regions, the left (with sites restricted to  $[1, L/2]$) and the right  (restricted to $[L/2+1,  L]$). The maximal entanglement entropy (EE) of this bubble configuration is 
\be 
S_{\rm max} (\alpha) = \sum_{k} \log m_1 (\alpha, k), 
\ee 
with $m_1 (\alpha, k)$ the Hilbert space dimension of bubble-left-region or bubble-right-region depending on which one is smaller,  
\be 
 m_1 (\alpha, k) = \left[ \sum_{\tilde{q} = 0} ^{q_k} D_{\rm therm } ( {\rm min }  (l_k ^{\rm left}, l_k ^{\rm right } ), \tilde{q}) \right], 
\ee 
with $l_k ^{\rm right} = l_k - l_k ^{\rm left} $. 
Correspondingly we introduce 
 $ 
m_2  (\alpha) =  \sum_{\tilde{q} = 0} ^{q_k} D_{\rm therm } ( {\rm max }  (l_k ^{\rm left}, l_k ^{\rm right } ), \tilde{q}). 
$ 
Assuming the eigenstates in the thermal bubbles are approximately random states, the Page-value~\cite{1993_Page_PRL} estimate of EE $S_{\rm pv} (\alpha)$ for this bubble configuration is given by 
\be 
S_{\rm pv} (\alpha) = \sum_{k} \left( \sum_{p = m_2+1} ^{m_1m_2} \frac{1}{p}  - \frac{m_2 - 1} {2m_1}  \right).  
\ee

Grouping the states with the same $\alpha$ together, the averaged EE (averaging over all eigenstates) can be 
rewritten as 
\be 
S_{\rm avg} = \sum_{\alpha} \left( \frac{D_H (\alpha)} {D_H} \right) S_{{\rm max}/{\rm pv}} (\alpha), 
\ee 
with $D_H (\alpha)$ the number of states having the same $\alpha$. In numerics, the weight $D_H(\alpha)/D_H$ can be easily sampled by randomly sampling Fock states $|\{n\}\rangle$ (with equal weight) because the probability follows 
\be 
P\left[ \alpha ( |\{n\}\rangle)  = \alpha_0 \right] = D_H(\alpha_0)/D_H. 
\ee

\section{Integrability of the thermal bubble for the nearest-neighbor-random-interaction model} 
In this section, we show that the thermal bubble for the particular model in Eq.(1) is exactly solvable by mapping to ``{\it inflated fermions}". For a given thermal-bubble many-body state, say $|1 0 0 1 001\rangle$, we can first add `$0$' in the front (the example state becomes $|\framebox{01} 0 \framebox{01} 0 \framebox{01}\rangle$, then the many-body state is made of `$\framebox{01}$'s and `$0$'s. We can group `$\framebox{01}$' together and make it an inflated fermion denoted as $\mathbb{1} \equiv \framebox{01}$. For the model in Eq.(1), we only have single-particle tunnelings in the thermal bubble state. The tunneling Hamiltonian in the inflated fermion basis is completely identical to that of the original fermions.  This can be proven by  considering tunneling processes one by one, as the coupling from $|\ldots \mathbb{1} 0 \ldots \rangle $ to $| \ldots 0 \mathbb{1} \ldots \rangle$  in the inflated-fermion basis maps to the coupling between  $|\ldots \framebox{0 1} 0 \ldots \rangle$ and $|\ldots 0 \framebox{0 1}\ldots \rangle$ in the original basis. This thermal bubble is then exactly solvable as the inflated fermions are non-interacting. The map also holds for hard core bosons.

Two remarks are in order. First, the inflated-fermion mapping restricts to models with homogenous tunnelings only.  An inhomogeneous term like  $h_j c_j ^\dag c_j$, induces a long-range string-like interaction between the inflated fermions, and the resulting model is no longer solvable. Second, the calculation of entanglement entropy using the inflated-fermion picture does not appear to be straightforward as the entanglement-cut may split one inflated fermion into two halves.

\section{The thermal phase} 
\begin{figure}[htp] 
\includegraphics[angle=0,width=\linewidth]{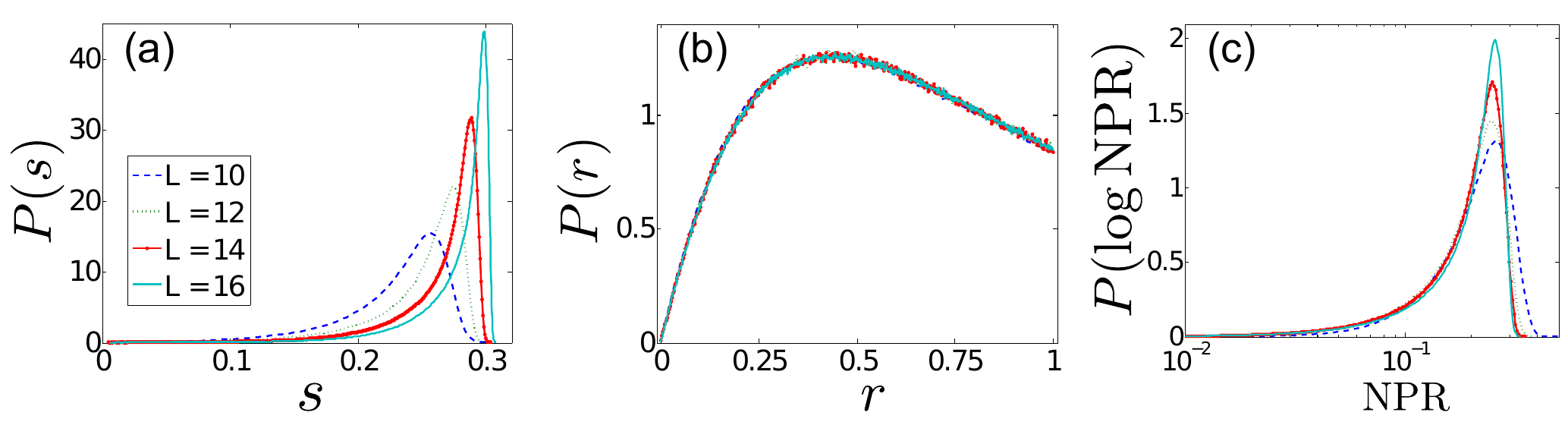}
 \caption{The quantum thermal phase in the random interaction model ({\it Eq.~(1) in the main text}).  
 (a), (b), and (c) show the probability distribution of half-chain entanglement entropy density ($P(s)$), the adjacent gap ratio ($ P(r)$), and the normalized participation ratio ($P( \log {\rm NPR})$), with a random interaction strength $W/t = 3$. The thermal phase obeys the GOE level statistics as shown by $P(r)$ in (b). As we increase the system size $L$, the distributions $P(s)$ and $P({\rm log NPR})$ get sharpen, implying the eigenstates are completely extended both in real space and in the Hilbert space at weak random interaction.  } 
 \label{fig:Fig3}  
\end{figure}

In this section, we give the results confirming thermalization of the random interaction model in the tunneling dominant regime.  The probability distributions of the different diagnostics in the thermal phase are shown in Fig.~\ref{fig:Fig3}. We see that the distributions for entanglement entropy and normalized participation ratio (NPR) develop sharp peaks at finite values of $s$ (entropy density) and $\log {\rm NPR}$, respectively (Fig.~\ref{fig:Fig3}(a,b)). This implies that the thermal phase is completely extended both in real space and in the many-body Hilbert space. Furthermore, the probability distribution of $r$-value collapses to the GOE form even for small system sizes with deviations barely noticeable as shown in Fig.~\ref{fig:Fig3}(b), providing strong numerical evidence for the many-body level repulsion in this model at weak random interaction.  All in all,  spinless fermions with random nearest neighbor interaction at weak randomness provide  one ideal model to investigate quantum thermalization, despite the translationally invariant interacting case being integrable.

\begin{figure}[htp] 
\includegraphics[angle=0,width=.8\linewidth]{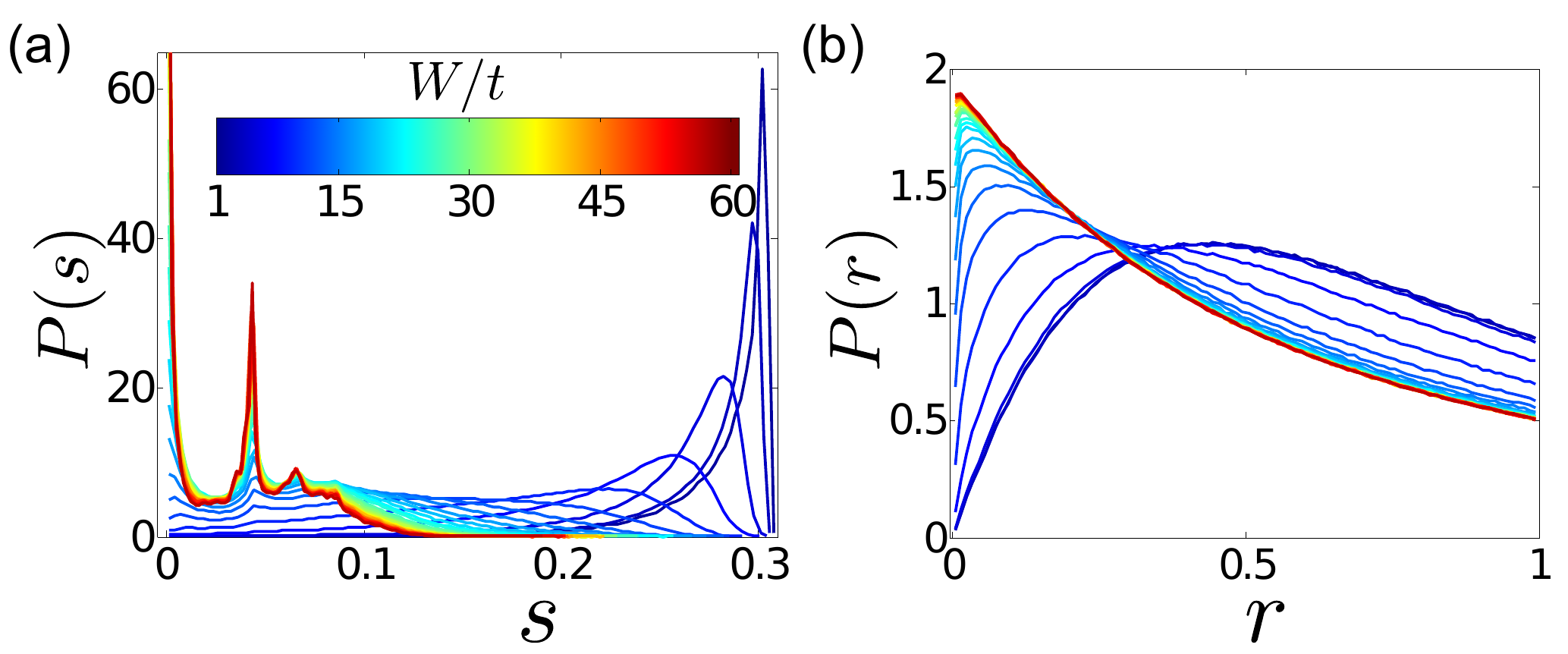}
 \caption{Cross-over from the thermal to MBL phases ($L = 16$). (a), and (b) show the distributions of the entanglement entropy density and gap ratio with varying $W/t$. In (b) we use the same color scheme to index $W/t$ as shown in (a).  } 
 \label{fig:Fig4}  
\end{figure}

\section{Cross-over from MBL to thermalization} 
In this section, we show the cross-over from MBL to thermalization. 
Fig.~\ref{fig:Fig4} shows how the probability distributions of entanglement entropy density and gap ration, $P(s)$ and $P(r)$, evolve in the crossover regime between thermal and localized 
phases. 
From Fig.~\ref{fig:Fig4}(a), we see that $s$ has a very broad 
distribution in the crossover regime. 
For $W/t > 15$, $P(s)$ has a strong peak at zero entanglement, indicating the dominance of insulating blocks.Upon decreasing $W/t$, the large-entanglement tail of $P(s)$ shifts rightward, corresponding to the increase in cross-block tunnelings. $P(r)$ is fairly robust at large $W/t$. As we decrease $W/t$, $P(r)$ quickly approaches the GOE distribution  once it starts to deviate from the Poisson case.  This strongly indicates GOE and Poisson distributions characterize two stable phases (thermal and MBL) in this model.

\section{MBL in the interacting Aubry-Andr\'e model} 
To show the MBL physics we present for random interactions is generic, we also provide the results for 
the Aubry-Andr\'e (AA) model.  On top of the original model  ({\it see Eq. (1) in the main text}), we now add an incommensurate potential, 
$$ 
\Delta H_{\rm AA}  =  2 \lambda \sum_j \cos (2\pi Q j) c_ j ^\dag c_j , 
$$  
with $Q$ an irrational number (here we use golden ratio $Q=\frac{1+\sqrt{5}}{2}$). As shown in Fig.~\ref{fig:AA}, we do not find any qualitative difference  from the pure random interaction case if the incommensurate potential is weak with single-particle Hamiltonian being extended.  It is worth noting here that for the AA model the thermal bubble is no-longer integrable.  We also mention that our numerical results (not shown) for the AA model in the localized single-particle case (i.e. $\lambda>t$ in contrast to Fig.~\ref{fig:AA} where $\lambda<t$ is considered explicitly) does not show any thermalization transition in the presence of random interactions implying that localized single-particle states remain localized when random interactions are turned on.

\begin{figure}[htp] 
\includegraphics[angle=0,width=\linewidth]{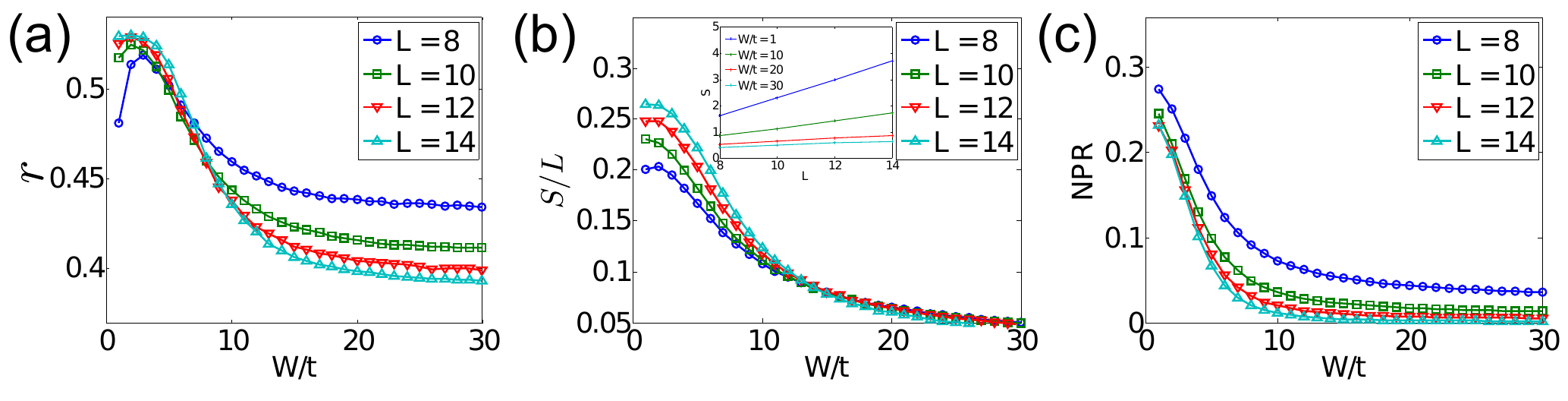}
 \caption{The MBL transition of the Aubry-Andr\'e model with random interaction. (a) The averaged gap ratio. (b) The entanglement entropy. Its inset shows the entanglement entropy scaling with increasing system size. (c) The normalized participation ratio. Here we choose $\lambda/t = 0.5$ where the single-particle AA Hamiltonian is extended. } 
 \label{fig:AA}  
\end{figure}

\section{MBL in the $(V+W)/(V-W)$ random interaction model} 
As a second model, we modify the original model by adding a constant interaction 
$$
\Delta H_{\rm int}  = V\sum_j n_j n_{j+1}. 
$$ 
Now the random interaction is drawn from $[V-W, V+W]$, instead of $[-W, W]$. 
As shown in Fig.~\ref{fig:Hint}, there is no qualitative difference from the the results presented in the main text.

\begin{figure}[htp] 
\includegraphics[angle=0,width=\linewidth]{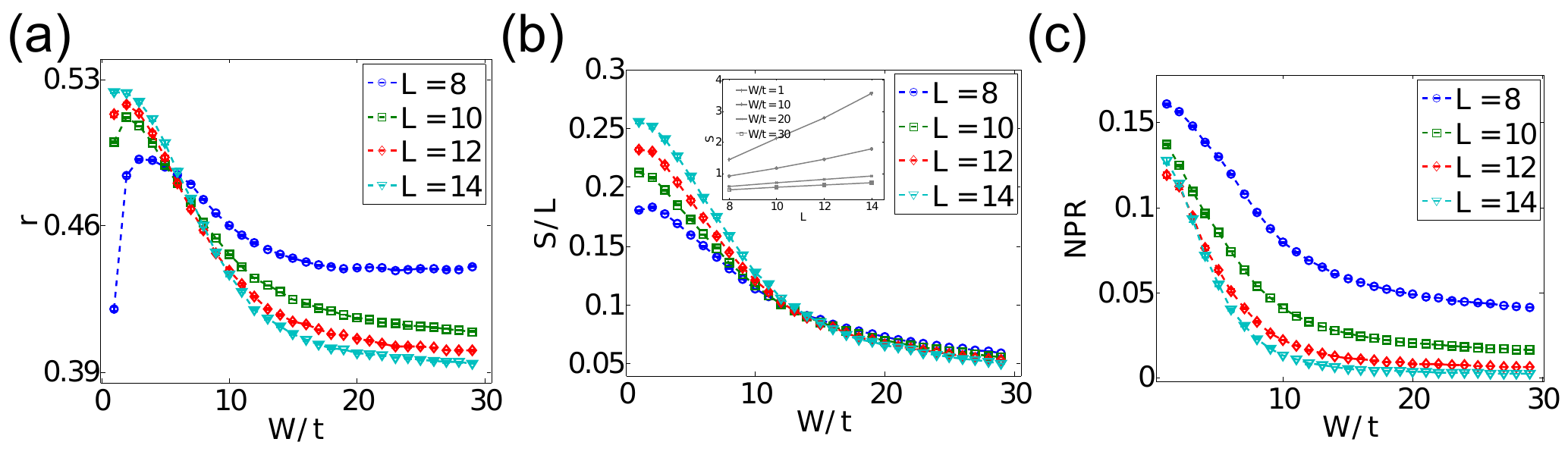}
 \caption{The MBL with random interaction with $V_j \in [V-W, V+W]$.  (a) The averaged gap ratio. (b) The entanglement entropy. Its inset shows the entanglement entropy scaling with increasing system size. (c) The normalized participation ratio.  Here we choose $V/t = 4$. } 
 \label{fig:Hint}  
\end{figure}

\end{widetext}

\bibliographystyle{apsrev4-1}
\bibliography{references}

\begin{thebibliography}{40}%
\makeatletter
\providecommand \@ifxundefined [1]{%
 \@ifx{#1\undefined}
}%
\providecommand \@ifnum [1]{%
 \ifnum #1\expandafter \@firstoftwo
 \else \expandafter \@secondoftwo
 \fi
}%
\providecommand \@ifx [1]{%
 \ifx #1\expandafter \@firstoftwo
 \else \expandafter \@secondoftwo
 \fi
}%
\providecommand \natexlab [1]{#1}%
\providecommand \enquote  [1]{``#1''}%
\providecommand \bibnamefont  [1]{#1}%
\providecommand \bibfnamefont [1]{#1}%
\providecommand \citenamefont [1]{#1}%
\providecommand \href@noop [0]{\@secondoftwo}%
\providecommand \href [0]{\begingroup \@sanitize@url \@href}%
\providecommand \@href[1]{\@@startlink{#1}\@@href}%
\providecommand \@@href[1]{\endgroup#1\@@endlink}%
\providecommand \@sanitize@url [0]{\catcode `\\12\catcode `\$12\catcode
  `\&12\catcode `\#12\catcode `\^12\catcode `\_12\catcode `\%12\relax}%
\providecommand \@@startlink[1]{}%
\providecommand \@@endlink[0]{}%
\providecommand \url  [0]{\begingroup\@sanitize@url \@url }%
\providecommand \@url [1]{\endgroup\@href {#1}{\urlprefix }}%
\providecommand \urlprefix  [0]{URL }%
\providecommand \Eprint [0]{\href }%
\providecommand \doibase [0]{http://dx.doi.org/}%
\providecommand \selectlanguage [0]{\@gobble}%
\providecommand \bibinfo  [0]{\@secondoftwo}%
\providecommand \bibfield  [0]{\@secondoftwo}%
\providecommand \translation [1]{[#1]}%
\providecommand \BibitemOpen [0]{}%
\providecommand \bibitemStop [0]{}%
\providecommand \bibitemNoStop [0]{.\EOS\space}%
\providecommand \EOS [0]{\spacefactor3000\relax}%
\providecommand \BibitemShut  [1]{\csname bibitem#1\endcsname}%
\let\auto@bib@innerbib\@empty
\bibitem [{\citenamefont {Anderson}(1958)}]{AndersonLocalization}%
  \BibitemOpen
  \bibfield  {author} {\bibinfo {author} {\bibfnamefont {P.~W.}\ \bibnamefont
  {Anderson}},\ }\href {\doibase 10.1103/PhysRev.109.1492} {\bibfield
  {journal} {\bibinfo  {journal} {Phys. Rev.}\ }\textbf {\bibinfo {volume}
  {109}},\ \bibinfo {pages} {1492} (\bibinfo {year} {1958})}\BibitemShut
  {NoStop}%
\bibitem [{\citenamefont {Basko}\ \emph {et~al.}(2006)\citenamefont {Basko},
  \citenamefont {Aleiner},\ and\ \citenamefont {Altshuler}}]{Basko06}%
  \BibitemOpen
  \bibfield  {author} {\bibinfo {author} {\bibfnamefont {D.}~\bibnamefont
  {Basko}}, \bibinfo {author} {\bibfnamefont {I.}~\bibnamefont {Aleiner}}, \
  and\ \bibinfo {author} {\bibfnamefont {B.}~\bibnamefont {Altshuler}},\ }\href
  {\doibase http://dx.doi.org/10.1016/j.aop.2005.11.014} {\bibfield  {journal}
  {\bibinfo  {journal} {Annals of Physics}\ }\textbf {\bibinfo {volume}
  {321}},\ \bibinfo {pages} {1126 } (\bibinfo {year} {2006})}\BibitemShut
  {NoStop}%
\bibitem [{\citenamefont {Nandkishore}\ and\ \citenamefont
  {Huse}(2015)}]{huse2015many}%
  \BibitemOpen
  \bibfield  {author} {\bibinfo {author} {\bibfnamefont {R.}~\bibnamefont
  {Nandkishore}}\ and\ \bibinfo {author} {\bibfnamefont {D.}~\bibnamefont
  {Huse}},\ }\href@noop {} {\bibfield  {journal} {\bibinfo  {journal} {Annu.
  Rev. Conden. Matter Phys.}\ }\textbf {\bibinfo {volume} {6}},\ \bibinfo
  {pages} {15} (\bibinfo {year} {2015})}\BibitemShut {NoStop}%
\bibitem [{\citenamefont {{Altman}}\ and\ \citenamefont
  {{Vosk}}(2015)}]{2015_Altman_review}%
  \BibitemOpen
  \bibfield  {author} {\bibinfo {author} {\bibfnamefont {E.}~\bibnamefont
  {{Altman}}}\ and\ \bibinfo {author} {\bibfnamefont {R.}~\bibnamefont
  {{Vosk}}},\ }\href {\doibase 10.1146/annurev-conmatphys-031214-014701}
  {\bibfield  {journal} {\bibinfo  {journal} {Annu. Rev. Condens. Matter
  Phys.}\ }\textbf {\bibinfo {volume} {6}},\ \bibinfo {pages} {383} (\bibinfo
  {year} {2015})}\BibitemShut {NoStop}%
\bibitem [{\citenamefont {Oganesyan}\ and\ \citenamefont
  {Huse}(2007)}]{oganesyan2007}%
  \BibitemOpen
  \bibfield  {author} {\bibinfo {author} {\bibfnamefont {V.}~\bibnamefont
  {Oganesyan}}\ and\ \bibinfo {author} {\bibfnamefont {D.~A.}\ \bibnamefont
  {Huse}},\ }\href@noop {} {\bibfield  {journal} {\bibinfo  {journal} {Phys.
  Rev. B}\ }\textbf {\bibinfo {volume} {75}},\ \bibinfo {pages} {155111}
  (\bibinfo {year} {2007})}\BibitemShut {NoStop}%
\bibitem [{\citenamefont {Pal}\ and\ \citenamefont {Huse}(2010)}]{pal2010}%
  \BibitemOpen
  \bibfield  {author} {\bibinfo {author} {\bibfnamefont {A.}~\bibnamefont
  {Pal}}\ and\ \bibinfo {author} {\bibfnamefont {D.~A.}\ \bibnamefont {Huse}},\
  }\href@noop {} {\bibfield  {journal} {\bibinfo  {journal} {Phys. Rev. B}\
  }\textbf {\bibinfo {volume} {82}},\ \bibinfo {pages} {174411} (\bibinfo
  {year} {2010})}\BibitemShut {NoStop}%
\bibitem [{\citenamefont {Bardarson}\ \emph {et~al.}(2012)\citenamefont
  {Bardarson}, \citenamefont {Pollmann},\ and\ \citenamefont
  {Moore}}]{moore2012}%
  \BibitemOpen
  \bibfield  {author} {\bibinfo {author} {\bibfnamefont {J.~H.}\ \bibnamefont
  {Bardarson}}, \bibinfo {author} {\bibfnamefont {F.}~\bibnamefont {Pollmann}},
  \ and\ \bibinfo {author} {\bibfnamefont {J.~E.}\ \bibnamefont {Moore}},\
  }\href {\doibase 10.1103/PhysRevLett.109.017202} {\bibfield  {journal}
  {\bibinfo  {journal} {Phys. Rev. Lett.}\ }\textbf {\bibinfo {volume} {109}},\
  \bibinfo {pages} {017202} (\bibinfo {year} {2012})}\BibitemShut {NoStop}%
\bibitem [{\citenamefont {Bauer}\ and\ \citenamefont
  {Nayak}(2013)}]{2013_Bauer_JSM}%
  \BibitemOpen
  \bibfield  {author} {\bibinfo {author} {\bibfnamefont {B.}~\bibnamefont
  {Bauer}}\ and\ \bibinfo {author} {\bibfnamefont {C.}~\bibnamefont {Nayak}},\
  }\href {http://stacks.iop.org/1742-5468/2013/i=09/a=P09005} {\bibfield
  {journal} {\bibinfo  {journal} {J. Stat. Mech. Theor. Exp.}\ }\textbf
  {\bibinfo {volume} {2013}},\ \bibinfo {pages} {P09005} (\bibinfo {year}
  {2013})}\BibitemShut {NoStop}%
\bibitem [{\citenamefont {Iyer}\ \emph {et~al.}(2013)\citenamefont {Iyer},
  \citenamefont {Oganesyan}, \citenamefont {Refael},\ and\ \citenamefont
  {Huse}}]{vadim}%
  \BibitemOpen
  \bibfield  {author} {\bibinfo {author} {\bibfnamefont {S.}~\bibnamefont
  {Iyer}}, \bibinfo {author} {\bibfnamefont {V.}~\bibnamefont {Oganesyan}},
  \bibinfo {author} {\bibfnamefont {G.}~\bibnamefont {Refael}}, \ and\ \bibinfo
  {author} {\bibfnamefont {D.~A.}\ \bibnamefont {Huse}},\ }\href {\doibase
  10.1103/PhysRevB.87.134202} {\bibfield  {journal} {\bibinfo  {journal} {Phys.
  Rev. B}\ }\textbf {\bibinfo {volume} {87}},\ \bibinfo {pages} {134202}
  (\bibinfo {year} {2013})}\BibitemShut {NoStop}%
\bibitem [{\citenamefont {Kj{\"a}ll}\ \emph {et~al.}(2014)\citenamefont
  {Kj{\"a}ll}, \citenamefont {Bardarson},\ and\ \citenamefont
  {Pollmann}}]{bardarson2014}%
  \BibitemOpen
  \bibfield  {author} {\bibinfo {author} {\bibfnamefont {J.~A.}\ \bibnamefont
  {Kj{\"a}ll}}, \bibinfo {author} {\bibfnamefont {J.~H.}\ \bibnamefont
  {Bardarson}}, \ and\ \bibinfo {author} {\bibfnamefont {F.}~\bibnamefont
  {Pollmann}},\ }\href@noop {} {\bibfield  {journal} {\bibinfo  {journal}
  {Phys. Rev. Lett.}\ }\textbf {\bibinfo {volume} {113}},\ \bibinfo {pages}
  {107204} (\bibinfo {year} {2014})}\BibitemShut {NoStop}%
\bibitem [{\citenamefont {Devakul}\ and\ \citenamefont
  {Singh}(2015)}]{2015_Singh_MBL_PRL}%
  \BibitemOpen
  \bibfield  {author} {\bibinfo {author} {\bibfnamefont {T.}~\bibnamefont
  {Devakul}}\ and\ \bibinfo {author} {\bibfnamefont {R.~R.~P.}\ \bibnamefont
  {Singh}},\ }\href {\doibase 10.1103/PhysRevLett.115.187201} {\bibfield
  {journal} {\bibinfo  {journal} {Phys. Rev. Lett.}\ }\textbf {\bibinfo
  {volume} {115}},\ \bibinfo {pages} {187201} (\bibinfo {year}
  {2015})}\BibitemShut {NoStop}%
\bibitem [{\citenamefont {Khemani}\ \emph {et~al.}(2016)\citenamefont
  {Khemani}, \citenamefont {Pollmann},\ and\ \citenamefont
  {Sondhi}}]{2015_Khemani_Pollmann_PRL}%
  \BibitemOpen
  \bibfield  {author} {\bibinfo {author} {\bibfnamefont {V.}~\bibnamefont
  {Khemani}}, \bibinfo {author} {\bibfnamefont {F.}~\bibnamefont {Pollmann}}, \
  and\ \bibinfo {author} {\bibfnamefont {S.~L.}\ \bibnamefont {Sondhi}},\
  }\href {\doibase 10.1103/PhysRevLett.116.247204} {\bibfield  {journal}
  {\bibinfo  {journal} {Phys. Rev. Lett.}\ }\textbf {\bibinfo {volume} {116}},\
  \bibinfo {pages} {247204} (\bibinfo {year} {2016})}\BibitemShut {NoStop}%
\bibitem [{\citenamefont {{Yu}}\ \emph {et~al.}(2015)\citenamefont {{Yu}},
  \citenamefont {{Pekker}},\ and\ \citenamefont
  {{Clark}}}]{2015_Yu_Pekker_arXiv}%
  \BibitemOpen
  \bibfield  {author} {\bibinfo {author} {\bibfnamefont {X.}~\bibnamefont
  {{Yu}}}, \bibinfo {author} {\bibfnamefont {D.}~\bibnamefont {{Pekker}}}, \
  and\ \bibinfo {author} {\bibfnamefont {B.~K.}\ \bibnamefont {{Clark}}},\
  }\href@noop {} {\bibfield  {journal} {\bibinfo  {journal} {ArXiv e-prints}\ }
  (\bibinfo {year} {2015})},\ \Eprint {http://arxiv.org/abs/1509.01244}
  {arXiv:1509.01244 [cond-mat.str-el]} \BibitemShut {NoStop}%
\bibitem [{\citenamefont {Lim}\ and\ \citenamefont
  {Sheng}(2016)}]{2015_Lim_Sheng_PRB}%
  \BibitemOpen
  \bibfield  {author} {\bibinfo {author} {\bibfnamefont {S.~P.}\ \bibnamefont
  {Lim}}\ and\ \bibinfo {author} {\bibfnamefont {D.~N.}\ \bibnamefont
  {Sheng}},\ }\href {\doibase 10.1103/PhysRevB.94.045111} {\bibfield  {journal}
  {\bibinfo  {journal} {Phys. Rev. B}\ }\textbf {\bibinfo {volume} {94}},\
  \bibinfo {pages} {045111} (\bibinfo {year} {2016})}\BibitemShut {NoStop}%
\bibitem [{\citenamefont {Kennes}\ and\ \citenamefont
  {Karrasch}(2016)}]{2016_Kennes_Karrasch_PRB}%
  \BibitemOpen
  \bibfield  {author} {\bibinfo {author} {\bibfnamefont {D.~M.}\ \bibnamefont
  {Kennes}}\ and\ \bibinfo {author} {\bibfnamefont {C.}~\bibnamefont
  {Karrasch}},\ }\href {\doibase 10.1103/PhysRevB.93.245129} {\bibfield
  {journal} {\bibinfo  {journal} {Phys. Rev. B}\ }\textbf {\bibinfo {volume}
  {93}},\ \bibinfo {pages} {245129} (\bibinfo {year} {2016})}\BibitemShut
  {NoStop}%
\bibitem [{\citenamefont {Imbrie}(2016)}]{imbrie2014many}%
  \BibitemOpen
  \bibfield  {author} {\bibinfo {author} {\bibfnamefont {J.~Z.}\ \bibnamefont
  {Imbrie}},\ }\href {\doibase 10.1007/s10955-016-1508-x} {\bibfield  {journal}
  {\bibinfo  {journal} {Journal of Statistical Physics}\ }\textbf {\bibinfo
  {volume} {163}},\ \bibinfo {pages} {998} (\bibinfo {year}
  {2016})}\BibitemShut {NoStop}%
\bibitem [{\citenamefont {Schreiber}\ \emph {et~al.}(2015)\citenamefont
  {Schreiber}, \citenamefont {Hodgman}, \citenamefont {Bordia}, \citenamefont
  {Lüschen}, \citenamefont {Fischer}, \citenamefont {Vosk}, \citenamefont
  {Altman}, \citenamefont {Schneider},\ and\ \citenamefont {Bloch}}]{blochmbl}%
  \BibitemOpen
  \bibfield  {author} {\bibinfo {author} {\bibfnamefont {M.}~\bibnamefont
  {Schreiber}}, \bibinfo {author} {\bibfnamefont {S.~S.}\ \bibnamefont
  {Hodgman}}, \bibinfo {author} {\bibfnamefont {P.}~\bibnamefont {Bordia}},
  \bibinfo {author} {\bibfnamefont {H.~P.}\ \bibnamefont {Lüschen}}, \bibinfo
  {author} {\bibfnamefont {M.~H.}\ \bibnamefont {Fischer}}, \bibinfo {author}
  {\bibfnamefont {R.}~\bibnamefont {Vosk}}, \bibinfo {author} {\bibfnamefont
  {E.}~\bibnamefont {Altman}}, \bibinfo {author} {\bibfnamefont
  {U.}~\bibnamefont {Schneider}}, \ and\ \bibinfo {author} {\bibfnamefont
  {I.}~\bibnamefont {Bloch}},\ }\href {\doibase 10.1126/science.aaa7432}
  {\bibfield  {journal} {\bibinfo  {journal} {Science}\ }\textbf {\bibinfo
  {volume} {349}},\ \bibinfo {pages} {842} (\bibinfo {year} {2015})},\ \Eprint
  {http://arxiv.org/abs/http://www.sciencemag.org/content/349/6250/842.full.pdf}
  {http://www.sciencemag.org/content/349/6250/842.full.pdf} \BibitemShut
  {NoStop}%
\bibitem [{\citenamefont {Bordia}\ \emph {et~al.}(2016)\citenamefont {Bordia},
  \citenamefont {L\"uschen}, \citenamefont {Hodgman}, \citenamefont
  {Schreiber}, \citenamefont {Bloch},\ and\ \citenamefont
  {Schneider}}]{2016_Bordia_Bloch_MBL_PRL}%
  \BibitemOpen
  \bibfield  {author} {\bibinfo {author} {\bibfnamefont {P.}~\bibnamefont
  {Bordia}}, \bibinfo {author} {\bibfnamefont {H.~P.}\ \bibnamefont
  {L\"uschen}}, \bibinfo {author} {\bibfnamefont {S.~S.}\ \bibnamefont
  {Hodgman}}, \bibinfo {author} {\bibfnamefont {M.}~\bibnamefont {Schreiber}},
  \bibinfo {author} {\bibfnamefont {I.}~\bibnamefont {Bloch}}, \ and\ \bibinfo
  {author} {\bibfnamefont {U.}~\bibnamefont {Schneider}},\ }\href {\doibase
  10.1103/PhysRevLett.116.140401} {\bibfield  {journal} {\bibinfo  {journal}
  {Phys. Rev. Lett.}\ }\textbf {\bibinfo {volume} {116}},\ \bibinfo {pages}
  {140401} (\bibinfo {year} {2016})}\BibitemShut {NoStop}%
\bibitem [{\citenamefont {{Bordia}}\ \emph {et~al.}(2016)\citenamefont
  {{Bordia}}, \citenamefont {{L{\"u}schen}}, \citenamefont {{Schneider}},
  \citenamefont {{Knap}},\ and\ \citenamefont
  {{Bloch}}}]{2016_Bordia_Bloch_arXiv}%
  \BibitemOpen
  \bibfield  {author} {\bibinfo {author} {\bibfnamefont {P.}~\bibnamefont
  {{Bordia}}}, \bibinfo {author} {\bibfnamefont {H.}~\bibnamefont
  {{L{\"u}schen}}}, \bibinfo {author} {\bibfnamefont {U.}~\bibnamefont
  {{Schneider}}}, \bibinfo {author} {\bibfnamefont {M.}~\bibnamefont {{Knap}}},
  \ and\ \bibinfo {author} {\bibfnamefont {I.}~\bibnamefont {{Bloch}}},\
  }\href@noop {} {\bibfield  {journal} {\bibinfo  {journal} {ArXiv e-prints}\ }
  (\bibinfo {year} {2016})},\ \Eprint {http://arxiv.org/abs/1607.07868}
  {arXiv:1607.07868 [cond-mat.quant-gas]} \BibitemShut {NoStop}%
\bibitem [{\citenamefont {yoon Choi1}\ \emph {et~al.}(2016)\citenamefont {yoon
  Choi1}, \citenamefont {Hild}, \citenamefont {Zeiher}, \citenamefont
  {Schauß}, \citenamefont {Rubio-Abadal}, \citenamefont {Yefsah},
  \citenamefont {Khemani}, \citenamefont {Huse}, \citenamefont {Bloch},\ and\
  \citenamefont {Gross1}}]{2016_Choi_Bloch_MBL_Science}%
  \BibitemOpen
  \bibfield  {author} {\bibinfo {author} {\bibfnamefont {J.}~\bibnamefont {yoon
  Choi1}}, \bibinfo {author} {\bibfnamefont {S.}~\bibnamefont {Hild}}, \bibinfo
  {author} {\bibfnamefont {J.}~\bibnamefont {Zeiher}}, \bibinfo {author}
  {\bibfnamefont {P.}~\bibnamefont {Schauß}}, \bibinfo {author} {\bibfnamefont
  {A.}~\bibnamefont {Rubio-Abadal}}, \bibinfo {author} {\bibfnamefont
  {T.}~\bibnamefont {Yefsah}}, \bibinfo {author} {\bibfnamefont
  {V.}~\bibnamefont {Khemani}}, \bibinfo {author} {\bibfnamefont {D.~A.}\
  \bibnamefont {Huse}}, \bibinfo {author} {\bibfnamefont {I.}~\bibnamefont
  {Bloch}}, \ and\ \bibinfo {author} {\bibfnamefont {C.}~\bibnamefont
  {Gross1}},\ }\href {\doibase 10.1126/science.aaf8834} {\bibfield  {journal}
  {\bibinfo  {journal} {Science}\ }\textbf {\bibinfo {volume} {352}},\ \bibinfo
  {pages} {1547} (\bibinfo {year} {2016})}\BibitemShut {NoStop}%
\bibitem [{\citenamefont {Smith}\ \emph {et~al.}(2016)\citenamefont {Smith},
  \citenamefont {Lee}, \citenamefont {Richerme}, \citenamefont {Neyenhuis},
  \citenamefont {Hess}, \citenamefont {Hauke}, \citenamefont {Heyl},
  \citenamefont {Huse},\ and\ \citenamefont
  {Monroe}}]{2015_Monroe_MBL_NatPhys}%
  \BibitemOpen
  \bibfield  {author} {\bibinfo {author} {\bibfnamefont {J.}~\bibnamefont
  {Smith}}, \bibinfo {author} {\bibfnamefont {A.}~\bibnamefont {Lee}}, \bibinfo
  {author} {\bibfnamefont {P.}~\bibnamefont {Richerme}}, \bibinfo {author}
  {\bibfnamefont {B.}~\bibnamefont {Neyenhuis}}, \bibinfo {author}
  {\bibfnamefont {P.~W.}\ \bibnamefont {Hess}}, \bibinfo {author}
  {\bibfnamefont {P.}~\bibnamefont {Hauke}}, \bibinfo {author} {\bibfnamefont
  {M.}~\bibnamefont {Heyl}}, \bibinfo {author} {\bibfnamefont {D.~A.}\
  \bibnamefont {Huse}}, \ and\ \bibinfo {author} {\bibfnamefont
  {C.}~\bibnamefont {Monroe}},\ }\href {\doibase 10.1038/nphys3783} {\bibfield
  {journal} {\bibinfo  {journal} {Nat. Phys.}\ } (\bibinfo {year} {2016}),\
  10.1038/nphys3783}\BibitemShut {NoStop}%
\bibitem [{\citenamefont {Huse}\ \emph {et~al.}(2014)\citenamefont {Huse},
  \citenamefont {Nandkishore},\ and\ \citenamefont
  {Oganesyan}}]{2014_Huse_MBL_PRB}%
  \BibitemOpen
  \bibfield  {author} {\bibinfo {author} {\bibfnamefont {D.~A.}\ \bibnamefont
  {Huse}}, \bibinfo {author} {\bibfnamefont {R.}~\bibnamefont {Nandkishore}}, \
  and\ \bibinfo {author} {\bibfnamefont {V.}~\bibnamefont {Oganesyan}},\ }\href
  {\doibase 10.1103/PhysRevB.90.174202} {\bibfield  {journal} {\bibinfo
  {journal} {Phys. Rev. B}\ }\textbf {\bibinfo {volume} {90}},\ \bibinfo
  {pages} {174202} (\bibinfo {year} {2014})}\BibitemShut {NoStop}%
\bibitem [{\citenamefont {Serbyn}\ \emph {et~al.}(2013)\citenamefont {Serbyn},
  \citenamefont {Papi\ifmmode~\acute{c}\else \'{c}\fi{}},\ and\ \citenamefont
  {Abanin}}]{2013_Serbyn_PRL}%
  \BibitemOpen
  \bibfield  {author} {\bibinfo {author} {\bibfnamefont {M.}~\bibnamefont
  {Serbyn}}, \bibinfo {author} {\bibfnamefont {Z.}~\bibnamefont
  {Papi\ifmmode~\acute{c}\else \'{c}\fi{}}}, \ and\ \bibinfo {author}
  {\bibfnamefont {D.~A.}\ \bibnamefont {Abanin}},\ }\href {\doibase
  10.1103/PhysRevLett.111.127201} {\bibfield  {journal} {\bibinfo  {journal}
  {Phys. Rev. Lett.}\ }\textbf {\bibinfo {volume} {111}},\ \bibinfo {pages}
  {127201} (\bibinfo {year} {2013})}\BibitemShut {NoStop}%
\bibitem [{\citenamefont {Chandran}\ \emph {et~al.}(2015)\citenamefont
  {Chandran}, \citenamefont {Kim}, \citenamefont {Vidal},\ and\ \citenamefont
  {Abanin}}]{chandran2015}%
  \BibitemOpen
  \bibfield  {author} {\bibinfo {author} {\bibfnamefont {A.}~\bibnamefont
  {Chandran}}, \bibinfo {author} {\bibfnamefont {I.~H.}\ \bibnamefont {Kim}},
  \bibinfo {author} {\bibfnamefont {G.}~\bibnamefont {Vidal}}, \ and\ \bibinfo
  {author} {\bibfnamefont {D.~A.}\ \bibnamefont {Abanin}},\ }\href {\doibase
  10.1103/PhysRevB.91.085425} {\bibfield  {journal} {\bibinfo  {journal} {Phys.
  Rev. B}\ }\textbf {\bibinfo {volume} {91}},\ \bibinfo {pages} {085425}
  (\bibinfo {year} {2015})}\BibitemShut {NoStop}%
\bibitem [{\citenamefont {Ros}\ \emph {et~al.}(2015)\citenamefont {Ros},
  \citenamefont {Müller},\ and\ \citenamefont {Scardicchio}}]{Ros2015420}%
  \BibitemOpen
  \bibfield  {author} {\bibinfo {author} {\bibfnamefont {V.}~\bibnamefont
  {Ros}}, \bibinfo {author} {\bibfnamefont {M.}~\bibnamefont {Müller}}, \ and\
  \bibinfo {author} {\bibfnamefont {A.}~\bibnamefont {Scardicchio}},\ }\href
  {\doibase http://dx.doi.org/10.1016/j.nuclphysb.2014.12.014} {\bibfield
  {journal} {\bibinfo  {journal} {Nuclear Physics B}\ }\textbf {\bibinfo
  {volume} {891}},\ \bibinfo {pages} {420 } (\bibinfo {year}
  {2015})}\BibitemShut {NoStop}%
\bibitem [{\citenamefont {Li}\ \emph {et~al.}(2015)\citenamefont {Li},
  \citenamefont {Ganeshan}, \citenamefont {Pixley},\ and\ \citenamefont
  {Das~Sarma}}]{2015_Li_MBL_PRL}%
  \BibitemOpen
  \bibfield  {author} {\bibinfo {author} {\bibfnamefont {X.}~\bibnamefont
  {Li}}, \bibinfo {author} {\bibfnamefont {S.}~\bibnamefont {Ganeshan}},
  \bibinfo {author} {\bibfnamefont {J.~H.}\ \bibnamefont {Pixley}}, \ and\
  \bibinfo {author} {\bibfnamefont {S.}~\bibnamefont {Das~Sarma}},\ }\href
  {\doibase 10.1103/PhysRevLett.115.186601} {\bibfield  {journal} {\bibinfo
  {journal} {Phys. Rev. Lett.}\ }\textbf {\bibinfo {volume} {115}},\ \bibinfo
  {pages} {186601} (\bibinfo {year} {2015})}\BibitemShut {NoStop}%
\bibitem [{\citenamefont {Modak}\ and\ \citenamefont
  {Mukerjee}(2015)}]{modak2015many}%
  \BibitemOpen
  \bibfield  {author} {\bibinfo {author} {\bibfnamefont {R.}~\bibnamefont
  {Modak}}\ and\ \bibinfo {author} {\bibfnamefont {S.}~\bibnamefont
  {Mukerjee}},\ }\href {\doibase 10.1103/PhysRevLett.115.230401} {\bibfield
  {journal} {\bibinfo  {journal} {Phys. Rev. Lett.}\ }\textbf {\bibinfo
  {volume} {115}},\ \bibinfo {pages} {230401} (\bibinfo {year}
  {2015})}\BibitemShut {NoStop}%
\bibitem [{\citenamefont {Li}\ \emph {et~al.}(2016)\citenamefont {Li},
  \citenamefont {Pixley}, \citenamefont {Deng}, \citenamefont {Ganeshan},\ and\
  \citenamefont {Das~Sarma}}]{2016_Li_Pixley_MBL_PRB}%
  \BibitemOpen
  \bibfield  {author} {\bibinfo {author} {\bibfnamefont {X.}~\bibnamefont
  {Li}}, \bibinfo {author} {\bibfnamefont {J.~H.}\ \bibnamefont {Pixley}},
  \bibinfo {author} {\bibfnamefont {D.-L.}\ \bibnamefont {Deng}}, \bibinfo
  {author} {\bibfnamefont {S.}~\bibnamefont {Ganeshan}}, \ and\ \bibinfo
  {author} {\bibfnamefont {S.}~\bibnamefont {Das~Sarma}},\ }\href {\doibase
  10.1103/PhysRevB.93.184204} {\bibfield  {journal} {\bibinfo  {journal} {Phys.
  Rev. B}\ }\textbf {\bibinfo {volume} {93}},\ \bibinfo {pages} {184204}
  (\bibinfo {year} {2016})}\BibitemShut {NoStop}%
\bibitem [{\citenamefont {{Hyatt}}\ \emph {et~al.}(2016)\citenamefont
  {{Hyatt}}, \citenamefont {{Garrison}}, \citenamefont {{Potter}},\ and\
  \citenamefont {{Bauer}}}]{2016_Bauer_arXiv}%
  \BibitemOpen
  \bibfield  {author} {\bibinfo {author} {\bibfnamefont {K.}~\bibnamefont
  {{Hyatt}}}, \bibinfo {author} {\bibfnamefont {J.~R.}\ \bibnamefont
  {{Garrison}}}, \bibinfo {author} {\bibfnamefont {A.~C.}\ \bibnamefont
  {{Potter}}}, \ and\ \bibinfo {author} {\bibfnamefont {B.}~\bibnamefont
  {{Bauer}}},\ }\href@noop {} {\bibfield  {journal} {\bibinfo  {journal} {ArXiv
  e-prints}\ } (\bibinfo {year} {2016})},\ \Eprint
  {http://arxiv.org/abs/1601.07184} {arXiv:1601.07184 [cond-mat.dis-nn]}
  \BibitemShut {NoStop}%
\bibitem [{\citenamefont {Vasseur}\ \emph {et~al.}(2016)\citenamefont
  {Vasseur}, \citenamefont {Friedman}, \citenamefont {Parameswaran},\ and\
  \citenamefont {Potter}}]{2016_Romain_Potter_PRB}%
  \BibitemOpen
  \bibfield  {author} {\bibinfo {author} {\bibfnamefont {R.}~\bibnamefont
  {Vasseur}}, \bibinfo {author} {\bibfnamefont {A.~J.}\ \bibnamefont
  {Friedman}}, \bibinfo {author} {\bibfnamefont {S.~A.}\ \bibnamefont
  {Parameswaran}}, \ and\ \bibinfo {author} {\bibfnamefont {A.~C.}\
  \bibnamefont {Potter}},\ }\href {\doibase 10.1103/PhysRevB.93.134207}
  {\bibfield  {journal} {\bibinfo  {journal} {Phys. Rev. B}\ }\textbf {\bibinfo
  {volume} {93}},\ \bibinfo {pages} {134207} (\bibinfo {year}
  {2016})}\BibitemShut {NoStop}%
\bibitem [{\citenamefont {{Bar Lev}}\ \emph {et~al.}(2016)\citenamefont {{Bar
  Lev}}, \citenamefont {{Reichman}},\ and\ \citenamefont
  {{Sagi}}}]{2016_Lev_arXiv}%
  \BibitemOpen
  \bibfield  {author} {\bibinfo {author} {\bibfnamefont {Y.}~\bibnamefont {{Bar
  Lev}}}, \bibinfo {author} {\bibfnamefont {D.~R.}\ \bibnamefont {{Reichman}}},
  \ and\ \bibinfo {author} {\bibfnamefont {Y.}~\bibnamefont {{Sagi}}},\
  }\href@noop {} {\bibfield  {journal} {\bibinfo  {journal} {ArXiv e-prints}\ }
  (\bibinfo {year} {2016})},\ \Eprint {http://arxiv.org/abs/1607.04652}
  {arXiv:1607.04652 [cond-mat.dis-nn]} \BibitemShut {NoStop}%
\bibitem [{cit()}]{citesupplement}%
  \BibitemOpen
  \href@noop {} {}\bibinfo {note} {See the Supplementary Material}\BibitemShut
  {NoStop}%
\bibitem [{\citenamefont {Page}(1993)}]{1993_Page_PRL}%
  \BibitemOpen
  \bibfield  {author} {\bibinfo {author} {\bibfnamefont {D.~N.}\ \bibnamefont
  {Page}},\ }\href {\doibase 10.1103/PhysRevLett.71.1291} {\bibfield  {journal}
  {\bibinfo  {journal} {Phys. Rev. Lett.}\ }\textbf {\bibinfo {volume} {71}},\
  \bibinfo {pages} {1291} (\bibinfo {year} {1993})}\BibitemShut {NoStop}%
\bibitem [{\citenamefont {{Khemani}}\ \emph {et~al.}(2016)\citenamefont
  {{Khemani}}, \citenamefont {{Lim}}, \citenamefont {{Sheng}},\ and\
  \citenamefont {{Huse}}}]{2016_Sheng_Huse_arXiv}%
  \BibitemOpen
  \bibfield  {author} {\bibinfo {author} {\bibfnamefont {V.}~\bibnamefont
  {{Khemani}}}, \bibinfo {author} {\bibfnamefont {S.~P.}\ \bibnamefont
  {{Lim}}}, \bibinfo {author} {\bibfnamefont {D.~N.}\ \bibnamefont {{Sheng}}},
  \ and\ \bibinfo {author} {\bibfnamefont {D.~A.}\ \bibnamefont {{Huse}}},\
  }\href@noop {} {\bibfield  {journal} {\bibinfo  {journal} {ArXiv e-prints}\ }
  (\bibinfo {year} {2016})},\ \Eprint {http://arxiv.org/abs/1607.05756}
  {arXiv:1607.05756 [cond-mat.dis-nn]} \BibitemShut {NoStop}%
\bibitem [{\citenamefont {Bell}(1972)}]{1972_Bell_RPP}%
  \BibitemOpen
  \bibfield  {author} {\bibinfo {author} {\bibfnamefont {R.~J.}\ \bibnamefont
  {Bell}},\ }\href {http://stacks.iop.org/0034-4885/35/i=3/a=306} {\bibfield
  {journal} {\bibinfo  {journal} {Reports on Progress in Physics}\ }\textbf
  {\bibinfo {volume} {35}},\ \bibinfo {pages} {1315} (\bibinfo {year}
  {1972})}\BibitemShut {NoStop}%
\bibitem [{\citenamefont {Wegner}(1980)}]{1980_Wegner_ZPB}%
  \BibitemOpen
  \bibfield  {author} {\bibinfo {author} {\bibfnamefont {F.}~\bibnamefont
  {Wegner}},\ }\href {\doibase 10.1007/BF01325284} {\bibfield  {journal}
  {\bibinfo  {journal} {Zeitschrift f{\"u}r Physik B Condensed Matter}\
  }\textbf {\bibinfo {volume} {36}},\ \bibinfo {pages} {209} (\bibinfo {year}
  {1980})}\BibitemShut {NoStop}%
\bibitem [{\citenamefont {Rodriguez}\ \emph {et~al.}(2011)\citenamefont
  {Rodriguez}, \citenamefont {Vasquez}, \citenamefont {Slevin},\ and\
  \citenamefont {R\"omer}}]{2011_Rodriguez_PRB}%
  \BibitemOpen
  \bibfield  {author} {\bibinfo {author} {\bibfnamefont {A.}~\bibnamefont
  {Rodriguez}}, \bibinfo {author} {\bibfnamefont {L.~J.}\ \bibnamefont
  {Vasquez}}, \bibinfo {author} {\bibfnamefont {K.}~\bibnamefont {Slevin}}, \
  and\ \bibinfo {author} {\bibfnamefont {R.~A.}\ \bibnamefont {R\"omer}},\
  }\href {\doibase 10.1103/PhysRevB.84.134209} {\bibfield  {journal} {\bibinfo
  {journal} {Phys. Rev. B}\ }\textbf {\bibinfo {volume} {84}},\ \bibinfo
  {pages} {134209} (\bibinfo {year} {2011})}\BibitemShut {NoStop}%
\bibitem [{\citenamefont {Luitz}\ \emph {et~al.}(2014)\citenamefont {Luitz},
  \citenamefont {Alet},\ and\ \citenamefont
  {Laflorencie}}]{2014_Luitz_Alet_PRL}%
  \BibitemOpen
  \bibfield  {author} {\bibinfo {author} {\bibfnamefont {D.~J.}\ \bibnamefont
  {Luitz}}, \bibinfo {author} {\bibfnamefont {F.}~\bibnamefont {Alet}}, \ and\
  \bibinfo {author} {\bibfnamefont {N.}~\bibnamefont {Laflorencie}},\ }\href
  {\doibase 10.1103/PhysRevLett.112.057203} {\bibfield  {journal} {\bibinfo
  {journal} {Phys. Rev. Lett.}\ }\textbf {\bibinfo {volume} {112}},\ \bibinfo
  {pages} {057203} (\bibinfo {year} {2014})}\BibitemShut {NoStop}%
\bibitem [{\citenamefont {Luitz}\ \emph {et~al.}(2015)\citenamefont {Luitz},
  \citenamefont {Laflorencie},\ and\ \citenamefont
  {Alet}}]{2015_Luitz_MBLedge_PRB}%
  \BibitemOpen
  \bibfield  {author} {\bibinfo {author} {\bibfnamefont {D.~J.}\ \bibnamefont
  {Luitz}}, \bibinfo {author} {\bibfnamefont {N.}~\bibnamefont {Laflorencie}},
  \ and\ \bibinfo {author} {\bibfnamefont {F.}~\bibnamefont {Alet}},\ }\href
  {\doibase 10.1103/PhysRevB.91.081103} {\bibfield  {journal} {\bibinfo
  {journal} {Phys. Rev. B}\ }\textbf {\bibinfo {volume} {91}},\ \bibinfo
  {pages} {081103} (\bibinfo {year} {2015})}\BibitemShut {NoStop}%
\bibitem [{\citenamefont {Vosk}\ \emph {et~al.}(2015)\citenamefont {Vosk},
  \citenamefont {Huse},\ and\ \citenamefont {Altman}}]{2015_Vosk_MBL_PRX}%
  \BibitemOpen
  \bibfield  {author} {\bibinfo {author} {\bibfnamefont {R.}~\bibnamefont
  {Vosk}}, \bibinfo {author} {\bibfnamefont {D.~A.}\ \bibnamefont {Huse}}, \
  and\ \bibinfo {author} {\bibfnamefont {E.}~\bibnamefont {Altman}},\ }\href
  {\doibase 10.1103/PhysRevX.5.031032} {\bibfield  {journal} {\bibinfo
  {journal} {Phys. Rev. X}\ }\textbf {\bibinfo {volume} {5}},\ \bibinfo {pages}
  {031032} (\bibinfo {year} {2015})}\BibitemShut {NoStop}%
\end{thebibliography}%

\end{document}